\documentclass[english]{article}
\usepackage[T1]{fontenc}
\usepackage[latin9]{inputenc}
\usepackage{amssymb}
\usepackage{esint}

\makeatletter
\newcommand{\lyxaddress}[1]{
\par {\raggedright #1
\vspace{1.4em}
\noindent\par}
}

\makeatother

\usepackage{babel}

\begin{document}

\title{Holography, charge and baryon asymmetry}

\author{T. R. Mongan}

\maketitle

\lyxaddress{84 Marin Avenue, Sausalito, CA 94965 USA, (415) - 332 - 1506, tmongan@gmail.com}

\begin{abstract}
The reason for baryon asymmetry in our universe has been a pertinent
question for many years. The holographic principle suggests a charged
preon model underlies the Standard Model of particle physics, and
any such charged preon model \emph{requires }baryon asymmetry. This
note estimates the baryon asymmetry predicted by charged preon models
in closed inflationary Friedmann universes. 
\end{abstract}
The reason for the dominance of matter over antimatter in our universe
has been a relevant issue for years \cite{key-1}. The holographic
principle \cite{key-2}, developed from black hole thermodynamics,
says all physics at a given point is described by the finite number
of bits of information on the particle horizon at the greatest distance
from which a light signal could reach the point since the end of inflation.
This suggests a charged preon model underlies the continuum mathematics
of Standard Model particle physics. There is a temperature associated
with the horizon and thermodynamics on the horizon implies gravity
is explained by Einstein\textquoteright{}s theory of general relativity
\cite{key-3}. This note shows that, in charged preon models, thermodynamics
on the horizon requires baryon asymmetry, and the baryon asymmetry
estimated for a closed universe is consistent with observations. This
simple explanation for baryon asymmetry suggests baryon asymmetry
and the resulting matter dominance in the universe are observational
evidence for a substructure beneath the Standard Model. It also suggests
the particle horizon is an appropriate focus for efforts to link gravity
with quantum mechanics.

The holographic principle says all information available about physics
\emph{within} a horizon at distance $d$ from an observer is given
by the finite amount of information \emph{on} the horizon. The number
of bits of information on the horizon, specified by one quarter of
the horizon area in Planck units \cite{key-2}, is $\pi d^{2}/(\delta^{2}\ln2)$.
The Planck length $\delta=\sqrt{\frac{\hbar G}{c^{3}}}$, where $G=6.67\times10^{-8}$
cm$^{3}$/g sec$^{2}$, $\hbar=1.05\times10^{-27}$g cm$^{2}$/sec,
and $c=3\times10^{10}$cm/sec. The following analysis relies on Bousso's
\cite{key-2} formulation of the holographic principle in terms of
the light sheets of the causal horizon, circumventing earlier objections
\cite{key-4} to using the holographic principle in cosmological contexts.
In particular, the argument applies to a vacuum-dominated closed universe,
created spontaneously by a quantum fluctuation, that can never collapse
\cite{key-5}

Because it involves continuum mathematics, the Standard Model can
only approximate an underlying finite-dimensional holographic theory.
In particular, a finite dimensional model involving only bits of information
on the horizon must describe all physics occurring within the horizon.
Linking bits of information on the horizon with Standard Model particles
requires a holographic model describing constituents (preons) of Standard
Model particles in terms of bits of information on the horizon.

All Standard Model particles have charges 0, 1/3, 2/3 or 1 in units
of the electron charge $\pm e$, so bits in a preon model must be
identified with fractional electric charge. Furthermore, in any physical
system, energy must be transferred to change information in a bit
from one state to another. Labelling the low energy state of a bit
$e/3n$ and the high energy state $-e/3n$ (where $n$ is some non-zero
integer depending on the particular preon model chosen) then amounts
to defining electric charge. If the universe is charge neutral (as
it must be if it began by a spontaneous quantum fluctuation from nothing)
there must be equal numbers of $e/3n$ and $-e/3n$ charges. A holographic
charged preon model in such a universe then embodies charge conservation,
a precondition for gauge invariance and Maxwell\textquoteright{}s
equations.

Protons have charge $e$ and anti-protons have charge $-e.$ Therefore,
regardless of the details of how bits of information on the horizon
specify a proton or anti-proton, the preon configuration specifying
a proton must differ in $3n$ bits from the configuration specifying
an anti-proton. Then, because $e/3n$ bits and $-e/3n$ bits do not
have the same energy, the number of protons and anti-protons created
in the early universe must be slightly different. In other words,
if $e/3n$ bits have lower energy than $-e/3n$ bits, there will inevitably
be more matter than anti-matter in the universe. However, a small
difference in energy of the bits on the horizon specifying a proton
or anti-proton is not inconsistent with protons and anti-protons having
identical mass.

The temperature at the time of baryon formation was $T_{B}=2m_{p}c^{2}/k=2.18\times10^{13}$
$^{o}K$, where the Boltzmann constant $k=1.38\times10^{-16}$(g cm$^{2}$/sec$^{2}$)/$^{o}K$,
and the proton mass $m_{p}=1.67\times10^{-14}$ g. So, the scale factor
of the universe at the time of baryogenesis was \cite{key-6} $R_{B}=R_{0}\left(\frac{2.725}{T_{B}}\right)\approx10^{15}$cm,
where 2.725 $^{o}K$ is today\textquoteright{}s cosmic microwave background
temperature and the scale factor of the universe today is $R_{0}\approx10^{28}$cm.
The time $t_{B}$ of baryogenesis, in seconds after the end of inflation,
can be determined from the Friedmann equation $\left(\frac{dR}{dt}\right)^{2}-\left(\frac{8\pi G}{3}\right)\varepsilon\left(\frac{R}{c}\right)^{2}=-\kappa c^{2}$.
After inflation, the universe is so large it is almost flat, so the
curvature parameter $\kappa\approx0$ . The energy density is $\varepsilon(R)=\varepsilon_{r}\left(\frac{R_{0}}{R}\right)^{4}+\varepsilon_{m}\left(\frac{R_{0}}{R}\right)^{3}+\varepsilon_{v}$
, where $\varepsilon_{r}$, $\varepsilon_{m}$ and $\varepsilon_{v}$
are, respectively, today's radiation, matter and vacuum energy densities.
Since the radiation energy density \cite{key-8} $\varepsilon_{r}=4\times10^{-13}$
erg/cm$^{3}$, the matter energy density $\varepsilon_{m}\approx9\times10^{-9}$
erg/cm$^{3}$, and vacuum energy density was negligible in the early
post-inflationary universe, the radiation term dominated when $R\ll10^{-5}R_{0}$,
before radiation/matter equality. Integrating $\left(\frac{dR}{dt}\right)^{2}-\left(\frac{8\pi G}{3c^{2}}\right)\frac{\varepsilon_{r}R_{0}^{4}}{R^{2}}=\left(\frac{dR}{dt}\right)^{2}-\frac{A^{2}}{R^{2}}=0$,
where $A=\sqrt{\frac{8\pi G\varepsilon_{r}R_{0}^{4}}{3c^{2}}}$ ,
from the end of inflation at $t=0$ to $t$ gives $\frac{1}{2}\left(R^{2}-R_{i}^{2}\right)=At$
, where $R_{i}$ is the scale factor of the universe at the end of
inflation. Therefore, $t_{B}=\frac{\left(R_{B}^{2}-R_{i}^{2}\right)}{2A}\approx\frac{R_{B}^{2}}{2A}\approx10^{-7}$seconds,
if $R_{B}\gg R_{i}$. The distance $d_{B}$ from any point in the
universe to the particle horizon for that point \cite{key-7} is $d_{B}=cR_{B}\intop_{0}^{t_{B}}\frac{dt'}{R\left(t'\right)}=\left[\frac{cR_{B}}{A}\sqrt{R_{i}^{2}+2At}\right]_{0}^{t_{B}}$
$=\frac{cR_{B}}{A}\left[\sqrt{R_{i}^{2}+2At_{B}}-R_{i}\right]$. Since
$R_{B}\gg R_{i}$, $d_{B}\approx cR_{B}\sqrt{\frac{2t_{B}}{A}}\approx10^{4}$cm.

The surface gravity on the particle horizon at baryogenesis is $g_{HB}=G\frac{4\pi}{3}\frac{\epsilon(R_{B})}{c^{2}}d_{B}\approx\frac{4\pi G}{3c}\epsilon_{r}\frac{R_{0}^{4}}{AR_{B}^{2}}$,
so the associated horizon temperature \cite{key-3} is $T_{HB}=\frac{\hbar}{2\pi ck}g_{HB}\approx6\times10^{-7}{}^{o}K$.
The temperature at any epoch is uniform throughout a post-inflationary
homogeneous isotropic Friedman universe, and the causal horizon at
baryogenesis is at distance $d_{B}$ from every point in the universe.
The temperature at every point on the causal horizon for every point
in the universe is the same because the surface gravity of the uniform
sphere within the horizon is the same at every point on every horizon.
The bits on all causal horizons are in thermal equilibrium, and there
are only two quantum states accessible to those bits. Therefore, the
use of equilibrium statistical mechanics is justified and the occupation
probabilities of the two bit states in thermal equilibrium at temperature
$T_{HB}$ are proportional to their corresponding Boltzmann factors.
So, if the energy of an $e/3n$ bit on the horizon at the time of
baryon formation is $E_{bit}-E_{d}$ and the energy of a $-e/3n$
bit is $E_{bit}+E_{d}$, the proton/antiproton ratio at baryogenesis
is $\left(\frac{e^{-\frac{E_{bit}-E_{d}}{kT_{HB}}}}{e^{-\frac{E_{bit}+E_{d}}{kT_{HB}}}}\right)^{3n}=e^{\frac{6nE_{d}}{kT_{HB}}}$
. Since $e^{\frac{6nE_{d}}{kT_{HB}}}\approx1+\frac{6nE_{d}}{kT_{HB}}$,
the proton excess is $\frac{6nE_{d}}{kT_{HB}}$. 

Any holographic preon model must link bits of information on the horizon
to bits of information specifying the location of preon constituents
of Standard Model particles within the universe. The wavefunction
specifying the probability distribution for the location of a particular
bit of information within the universe has only two energy levels.
The energy released when a bit in the universe drops from the (1)
to the (0) state raises another bit from the (0) to the (1) state,
and that is the mechanism for charge conservation. The energy must
be transferred by a massless quantum with wavelength related to the
size of the universe. There is no reliable definition of the size
(as opposed to the scale factor) of a flat or open universe, so it
is necessary to restrict the analysis to closed Friedmann universes.
The only macroscopic length characteristic of the size of a closed
Friedmann universe with radius (scale factor) $R\left(t\right)$ is
the circumference $2\pi R\left(t\right)$. If the energy $2E_{d}$
to change the state of a bit associated with a preon within the universe
(and the corresponding bit on the horizon) at baryogenesis equals
the energy of massless quanta with wavelength characteristic of the
size of a closed Friedmann universe with radius $R_{B}$, $2E_{d}=\frac{\hbar c}{R_{B}}$
. Then, substituting from above, the proton excess at baryogenesis
is $\frac{6nE_{d}}{kT_{HB}}=\left(\frac{12n\pi c^{2}}{R_{0}}\right)\left(\frac{2.725}{T_{B}}\right)\sqrt{\frac{3}{8\pi G\varepsilon_{r}}}$.
The dependence on $R_{0}$ arises because $R_{B}$, the radius of
the universe at baryogenesis, depends on $R_{0}$, today\textquoteright{}s
cosmic microwave background temperature 2.725 $^{o}K$, and the temperature
$T_{B}$ at baryogenesis. For $R_{0}\approx10^{28}$cm, the proton
excess is $\frac{6nE_{d}}{kT_{HB}}\approx0.9n\times10^{-9}$.

The WMAP estimate \cite{key-9} of baryon density to cosmic microwave
background photon density ratio is $6.1\times10^{-10}$. A charged
preon model \cite{key-10} with $n=2$ involves three strands, with
charged bits on the end of each strand, bound by non-local forces
into each Standard Model particle. At the time of baryogenesis, the
nnmber of proton states with six $e/6$ bits, the number of anti-protons
states with six $-e/6$ bits, and the number of photon states with
three $e/6$ and three $-e/6$ bits are approximately equal. Then,
when almost all protons and anti-protons annihilate to two photons,
the ratio of baryon to photon states is $\frac{1}{3}(1.8\times10^{-9})=6\times10^{-10}$,
in good agreement with the WMAP result.

If $R_{0}\approx10^{28}$cm, the model in this paper also predicts
a positron excess of $\approx1.7n\times10^{-6}$ when the universe
cools to the point where electron-positron pairs can survive. This
primordial positron excess is a primary source of positrons that might
help explain cosmic ray positron excess in the PAMELA experiment \cite{key-11}.
The positron excess might also explain part of the asymmetric 511
keV gamma radiation from the galactic center \cite{key-12}.

\end{document}